\documentstyle[11pt]{article}

\columnwidth\textwidth

\begin{document} %\draft

\title{General massive gauge theory}

\author{G\"unter Scharf\\ Institut f\"ur Theoretische Physik der Universit\"at
Z\"urich\\ Winterthurerstrasse 190, 8057 Z\"urich, Switzerland} \date{}

\maketitle 
\advance\baselineskip by 10 pt \begin{abstract} The concept of
perturbative gauge invariance formulated exclusively by means of asymptotic
fields is used to construct massive gauge theories. We consider the interactions
of $r$ massive and $s$ massless gauge fields together with $(r+s)$ fermionic
ghost and anti-ghost fields. First order gauge invariance requires the
introduction of unphysical scalars (Goldstone bosons) and fixes their trilinear
couplings. At second order additional physical scalars (Higgs fields) are
necessary, their coupling is further restricted at third order.  In case of one
physical scalar all couplings are determined by gauge invariance, including the
Higgs potential. For three massive and one massless gauge field the $SU(2)\times
U(1)$ electroweak theory comes out as the unique solution.

properties of perturbation theory \vskip 1cm Keywords: Massive gauge theories,
electroweak theory \vskip 1cm

\end{abstract} \newpage %\pacs{11.15.-q} % 11.15.-q Gauge field theories

%\narrowtext

\def\fii{\varphi} \def\d{\partial}\def\=d{\,{\buildrel\rm def\over =}\,}
\def\dh{\mathop{\vphantom{\odot}\hbox{$\partial$}}} \def\dl{\dh^\leftrightarrow}
\def\sqr#1#2{{\vcenter{\vbox{\hrule height.#2pt\hbox{\vrule width.#2pt
height#1pt \kern#1pt \vrule width.#2pt}\hrule height.#2pt}}}}
\def\w{\mathchoice\sqr45\sqr45\sqr{2.1}3\sqr{1.5}3\,} \def\eps{\varepsilon}
\def\oe{\overline{\rm e}} \def\onu{\overline{\nu}}
\def\ds{\hbox{\rlap/$\partial$}} \def\lra{\longrightarrow}

\section{Introduction} In gauge theories with massive gauge bosons the masses
are conventionally generated by the Higgs mechanism \cite{1}.  One introduces
scalar fields into the theory which have asymmetric self-interactions so that
some physical scalar field gets a symmetry breaking vacuum expectation value
(Higgs field).  Then the gauge symmetry is spontaneously broken and the gauge
fields can acquire mass. Here the notion "gauge symmetry" refers to the symmetry
of the total action.

For various reasons there are still considerable doubts whether the above
picture is really fundamental, one being the ad-hoc character of the
construction. However, it is possible to consider massive gauge theories from a
quite different point of view. If one takes the adiabatically switched S-matrix
$S(g)$  ($g(x)$ a Schwartz test function) as the basic object, defined by the
perturbation series \cite{2} $$S(g)=1+\sum_{n=1}^\infty{1\over n!}\int
d^4x_1\ldots d^4x_n\, T_n(x_1,\ldots ,x_n)g(x_1)\ldots g(x_n),\eqno(1.1)$$ then
one would like to formulate gauge invariance in terms of the time-ordered
products $T_n$. Since the latter are expressed by the asymptotic free fields, it
is a priori not clear whether such a perturbative definition of gauge invariance
is possible. We have found that this is indeed the case \cite{3}, no matter if
the gauge fields are massless or massive \cite{4}. The definition of
perturbative gauge invariance reads as follows $$d_QT_n\=d
[Q,T_n]=i\sum_{l=1}^n{\d\over\d x_l^\mu}T^\mu_{n/l} (x_1,\ldots x_l\ldots
x_n).\eqno(1.2)$$ Here $Q$ is the nilpotent gauge charge, first introduced by
Kugo and Ojima \cite{5}, and the $T^\nu_{n/l}$ are time-ordered products with a
so-called $Q$-vertex at $x_l$. These quantities are defined in the next section
and in sect.3.

The idea of the paper is to start from a general ansatz for $T_1(x)$ and to use
perturbative gauge invariance (1.2) to determine the coupling parameters in
$T_1$. This is a straightforward generalization of \cite{4} with the merit that
in the more general framework the discussion is simpler and more transparent.
The general ansatz contains massless and massive gauge fields and ghosts, as
well as unphysical (Goldstone bosons) and physical scalar (Higgs) fields. In
contrast to standard theory of spontaneous symmetry breaking where the scalar
fields are members of some multiplet, we treat the unphysical and physical
scalars completely free and independent. This turns out to be natural because
their couplings come out quite different: the coupling of the unphysical scalars
is (up to mass dependent factors) given by the structure constants $f_{abc}$ of
the gauge group Lie algebra (sect.3), whereas the Higgs couplings are of a
different diagonal type (sect.4). Nevertheless, in the case of one physical
scalar the resulting couplings are in agreement with the usual theory, including
the asymmetric Higgs potential (sect.5). For more than one Higgs field their
couplings are not completely determined by gauge invariance.

As a consequence of perturbative gauge invariance we find many relations between
the masses of the gauge fields and the structure constants $f_{abc}$. As an
application we consider in sect.6 the physical case of three massive gauge
fields and one massless (photon) field and ask the question: what are the
possible gauge theories ? The relations of gauge invariance enables us to
calculate the $f_{abc}$ in terms of the masses. The unique result is the usual
$SU(2)\times U(1)$ electroweak theory. In this way the standard theory looses
its ad-hoc character.

The same problem has recently been considered by D.R. Grigore \cite{18} using a
different definition of gauge invariance. Most of his results are in agreement
with ours, only his treatment of the Higgs fields is misleading.

\section{A general massive gauge theory}

We consider $r$ massive and $s$ massless gauge fields $A_a^\mu$, $a=1,\ldots
,r+s$ together with $(r+s)$ fermionic ghost and anti-ghost fields $u_a, \tilde
u_a$. These free asymtotic fields are quantized as follows $$(\w
+m_a^2)A_a^\mu(x)=0,\>[A_a^\mu(x),A_b^\nu(y)]_-=i\delta_{ab}g^
{\mu\nu}D_{m_a}(x-y),\eqno(2.1)$$ $$(\w +m_a^2)u_a(x)=0=(\w +m_a^2)\tilde
u_a(x)\eqno(2.2)$$ $$\{u_a(x),\tilde
u_b(y)\}_+=-i\delta_{ab}D_{m_a}(x-y),\eqno(2.3)$$ all other commutators vanish,
$D_m$ are the Jordan-Pauli distributions. The masses of a gauge field and the
corresponding ghost and anti-ghost fields must be equal, otherwise perturbative
gauge invariance cannot be achieved. We have $m_a=0$ for $a>r$.

In order to get a gauge charge $Q$ which is nilpotent $$Q^2=0,\eqno(2.4)$$ we
have to introduce for every massive gauge vector field $A_a^\mu(x), a\le r,$ a
scalar partner $\Phi_a(x)$ with the same mass $m_a$. The scalar fields are
quantized according to $$(\w
+m_a^2)\Phi_a(x)=0,\>[\Phi_a(x),\Phi_b(y)]=-i\delta_{ab}
D_{m_a}(x-y).\eqno(2.5)$$ Then the gauge charge $Q$ is defined by $$Q\=d \int
d^3x\,(\d_{\nu}A_a^{\nu}+m_a\Phi_a){\dl}_0u_a.  \eqno(2.6)$$ Calculating $Q^2$
as one half of the anticommutator $\{Q,Q\}$ one easily verifies the nilpotency
(2.4).

The scalar and ghost fields appearing in $Q$ (2.6) are all unphysical because
their excitations do not belong to the physical subspace \cite{7} $${\cal
H}_{\rm phys}={\rm Ker}Q\Bigl/{\rm Ran}Q.\eqno(2.7)$$ To discuss this in detail
it is necessary to introduce a concrete representation of the various asymptotic
fields in Fock space.  We want to avoid that to stress the fact that our
definition of gauge invariance refers to a structural property independent of
representation. Then we simply call a field unphysical if it appears in $Q$
(2.6), otherwise it is physical. For the gauge fields that means $\d_\nu A^\nu$
is unphysical.  Second order gauge invariance will force us to introduce
additional $physical$ scalar fields $\fii_p, p=1,\ldots ,t$, called Higgs
fields, with arbitrary masses $\mu_p$. We shall use indices $p,q,\ldots=1,\ldots
t$ from the end of the alphabet to number the Higgs fields, letters
$h,j,k,l,\ldots=1,\ldots r$ from the middle denote the other massive fields and
$a,b,c,d,e,f,\ldots=1,\ldots r+s$ is used for unrestricted 'color' indices.

With this field content we are going to analyse the following trilinear
couplings: $$T_1(x)=T_1^0+T_1^1+\ldots +T_1^{11}\eqno(2.8)$$ where
$$T_1^0=igf_{abc}(A_{\mu a}A_{\nu b}\d^\nu A_c^\mu-A_{\mu a}u_b\d^\mu \tilde
u_c)\eqno(2.9)$$
$$T_1^1=igf^1_{ahj}A_a^\mu(\Phi_h\d_\mu\Phi_j-\Phi_j\d_\mu\Phi_h), \quad
f^1_{ahj}=-f^1_{ajh}\eqno(2.10)$$ $$T_1^2=igf^2_{abh}A_{\mu
a}A^\mu_b\Phi_h,\quad f^2_{abh}=f^2_{bah} \eqno(2.11)$$
$$T_1^3=igf^3_{abh}\tilde u_au_b\Phi_h\eqno(2.12)$$
$$T_1^4=igf^4_{hjk}\Phi_h\Phi_j\Phi_k,\eqno(2.13)$$ where $f^4_{hjk}$ is totally
symmetric in $h,j,k$ and $g$ is a coupling constant. All $f$'s are real because
$T_1$ must be skew-adjoint. For reasons of economy we assume the pure Yang-Mills
coupling $f_{abc}$ in (2.9) to be totally antisymmetric. If one starts with the
most general ansatz, one must repeat the discussion in \cite{21} to derive the
antisymmetry. The Jacobi identity need not be assumed, it follows explicitly
below in second order (Sect.4.1). In $T_1^1$ we have only considered the
antisymmetric combination because the symmetric one can be expressed by a
divergence $$A^\mu_a(\Phi_h\d_\mu\Phi_j+\Phi_j\d_\mu\Phi_h)=\d_\mu(A_a^\mu\Phi_h
\Phi_j)-\d_\mu A_a^\mu\Phi_h\Phi_j.$$ The remaining $\d_\mu A_a^\mu$ term is a
coboundary $d_Q(\tilde u_a \Phi_h\Phi_j)$ plus terms of the form $T_1^3, T_1^4$.
But divergence and coboundary couplings can always be skipped in the discussion
of perturbative gauge invariance \cite{12}.

The Higgs couplings are obtained by replacing the scalar fields in (2.10-13) by
Higgs fields: $$T_1^5=igf^5_{ahp}A_a^\mu(\Phi_h\d_\mu\fii_p-\fii_p\d_\mu\Phi_h)
\eqno(2.14)$$ $$T_1^6=igf^6_{apq}A_a^\mu(\fii_p\d_\mu\fii_q-\fii_q\d_\mu\fii_p),
\quad f^6_{apq}=-f^6_{aqp}\eqno(2.15)$$ $$T_1^7=igf^7_{abp}A_{\mu
a}A_b^\mu\fii_p,\quad f^7_{abp}=f^7_{bap} \eqno(2.16)$$
$$T_1^8=igf^8_{abp}\tilde u_au_b\fii_p\eqno(2.17)$$
$$T_1^9=igf^9_{hjp}\Phi_h\Phi_j\fii_p,\quad f^9_{hjp}=f^9_{jhp} \eqno(2.18)$$
$$T_1^{10}=igf^{10}_{hpq}\Phi_h\fii_p\fii_q,\quad f^{10}_{hpq}=f^{10}
_{hqp}\eqno(2.19)$$ $$T_1^{11}=igf^{11}_{pqu}\fii_p\fii_q\fii_u, \eqno(2.20)$$
where $f^{11}$ is totally symmetric. {\it All products of field operators
throughout are normally ordered (Wick) products of free fields.} Interacting
fields do not appear at all.

\section{First order gauge invariance}

The gauge charge $Q$ (2.6) defines a gauge variation according to $$d_Q F\=d
QF-(-1)^{n_F}FQ,\eqno(3.1)$$ where $n_F$ is the number of ghost plus anti-ghost
fields in the Wick monomial $F$. We get the following gauge variations of the
fundamental fields $$d_QA_a^\mu(x)=i\d^\mu u_a(x),\quad d_Q\Phi_h(x)=im_h
u_h(x)\eqno(3.2)$$ $$d_Qu_a(x)=0,\quad d_Q\tilde u_a(x)=-i(\d_\mu
A_a^\mu(x)+m_a\Phi_a(x)) \eqno(3.3)$$ $$d_Q\fii_p=0.\eqno(3.4)$$ These
infinitesimal gauge transformations have some similarity with the BRST
transformations \cite{11}, but we emphasize the following differences. The BRST
transformations are defined for interacting fields, whereas we work with
asymptotic free fields only and establish gauge invariance order by order. BRST
invariance only holds if the quadratic free Lagrangian, the gauge fixing term
and the quartic term in the action are also transformed. We have no such terms
in $T_1$ (2.8) so that the compensations of terms in the gauge variations are
totally different.

We now calculate the gauge variation of all terms in $T_1$ and transform the
result to a divergence form $$d_QT_1=i\d_\mu T^\mu_{1/1}.\eqno(3.5)$$ The
$T^\mu_{1/1}$ appearing here is the $Q$-vertex. It is not unique, but the
possible modification has no influence on gauge invariance of higher orders
\cite{12}. The most convenient way to achieve the divergence form (3.5) is to
take out the derivatives of the ghost fields and use the field equations. In
this way we find: $$d_QT_1^0=gf_{abc}\Bigl\{\d_\mu[A_{\nu a}u_b(\d^\nu
A_c^\mu-\d^\mu A_c^\nu) +{1\over 2}u_au_b\d^\mu\tilde u_c]$$ $$-m_c^2A_{\nu
a}u_bA_c^\nu+{1\over 2}m_c^2u_au_b\tilde u_c+m_cA_{\nu a}
u_b\d^\nu\Phi_c\Bigl\}\eqno(3.6)$$
$$d_QT_1^1=-gf^1_{ahj}\Bigl\{\d^\mu[u_a(\Phi_h\d_\mu\Phi_j-\Phi_j\d_
\mu\Phi_h)$$
$$+m_jA_a^\mu\Phi_hu_j-m_hA_a^\mu\Phi_ju_h]+(m_j^2-m_h^2)u_a\Phi_h\Phi_j$$
$$+m_h(\d_\mu A_a^\mu\Phi_j+2A_a^\mu\d_\mu\Phi_j)u_h-m_j(\d_\mu A_a^\mu
\Phi_h+2A_a^\mu\d_\mu\Phi_h)u_j\Bigl\}\eqno(3.7)$$
$$d_QT_1^2=-gf^2_{abh}\Bigl\{\d_\mu(u_aA_b^\mu+A_a^\mu u_b)\Phi_h -u_a\d_\mu
A_b^\mu\Phi_h-u_aA_b^\mu\d_\mu\Phi_h$$ $$-u_b\d_\mu
A_a^\mu\Phi_h-u_bA_a^\mu\d_\mu\Phi_h+m_hA_{\mu a} A_b^\mu u_h\Bigl\}\eqno(3.8)$$
$$d_QT_1^3=gf^3_{abh}\Bigl\{(\d_\mu A_a^\mu+m_a\Phi_a)u_b\Phi_h- m_h\tilde
u_au_bu_h\Bigl\}\eqno(3.9)$$
$$d_QT_1^4=-gf^4_{hjk}\Bigl\{m_hu_h\Phi_j\Phi_k+m_j\Phi_hu_j\Phi_k
+m_k\Phi_h\Phi_ju_k\Bigl\}\eqno(3.10)$$
$$d_QT_1^5=-gf^5_{ahp}\Bigl\{\d^\mu[u_a(\Phi_h\d_\mu\fii_p-
\fii_p\d_\mu\Phi_h)-m_hA_a^\mu\fii_pu_h]$$
$$+(m_h^2-\mu_p^2)u_a\Phi_h\fii_p+2m_hA_a^\mu u_h\d_\mu\fii_p +m_h\d_\mu A_a^\mu
u_h\fii_p\Bigl\}\eqno(3.11)$$
$$d_QT_1^6=-gf^6_{apq}\Bigl\{\d^\mu[u_a(\fii_p\d_\mu\fii_q- \fii_q\d_\mu\fii_p)]
+(\mu_q^2-\mu_p^2)u_a\fii_p\fii_q\Bigl\}\eqno(3.12)$$
$$d_QT_1^7=-gf^7_{abp}\Bigl\{\d^\mu[(u_aA_{\mu b}+u_bA_{\mu a})\fii_p] $$
$$-(u_a\d_\mu A_b^\mu+u_b\d_\mu A_a^\mu)\fii_p-(u_aA_b^\mu+
u_bA_a^\mu)\d_\mu\fii_p\Bigl\}\eqno(3.13)$$ $$d_qT_1^8=gf^8_{abp}(\d_\mu
A_a^\mu+m_a\Phi_a)u_b\fii_p\eqno(3.14)$$
$$d_QT_1^9=-gf^9_{hjp}(m_hu_h\Phi_j+m_ju_j\Phi_h)\fii_p\eqno(3.15)$$
$$d_QT_1^{10}=-gf^{10}_{hpq}m_hu_h\fii_p\fii_q.\eqno(3.16)$$

We have given this long list in detail because a lot of information can directly
be read off. The divergence terms give the Q-vertex
$$T_{1/1}^\mu=gf_{abc}\Bigl[A_{\nu a}u_b(\d^\nu A_c^\mu-\d^\mu A_c^\nu) +{1\over
2}u_au_b\d^\mu\tilde u_c\Bigl]\eqno(3.17.1)$$
$$-gf^1_{ahj}\Bigl[2u_a\Phi_h\d^\mu\Phi_j+m_jA_a^\mu\Phi_hu_j-
m_hA_a^\mu\Phi_ju_h\Bigl]\eqno(3.17.2)$$
$$-gf^2_{abh}(u_aA_b^\mu+u_bA_a^\mu)\Phi_h\eqno(3.17.3)$$
$$-gf^5_{ahp}\Bigl[u_a(\Phi_h\d^\mu\fii_p-\fii_p\d^\mu\Phi_h) -m_hA_a^\mu
u_h\fii_p\Bigl]\eqno(3.17.4)$$
$$-2gf^6_{apq}u_a\fii_p\d^\mu\fii_q\eqno(3.17.5)$$
$$-gf^7_{abp}(u_aA_b^\mu+u_bA_a^\mu)\fii_p.\eqno(3.17.6)$$

The remaining terms must cancel out. Collecting the terms $\sim u_b A_{\mu
a}A_c^\mu$ we get the relation
$$2m_bf^2_{acb}=(m_a^2-m_c^2)f_{abc}.\eqno(3.18)$$ Hence, if $m_b=0$ and
$f_{abc}\ne 0$ we must have $$m_a=m_c.\eqno(3.19)$$ For $m_b, m_h\ne 0$ we find
$$f^2_{abh}={m_b^2-m_a^2\over 2m_h}f_{abh}.\eqno(3.20)$$ Then, collecting terms
$\sim A_{\mu a}u_h\d^\mu\Phi_j$ we get $$f^1_{ahj}={m_j^2+m_h^2-m_a^2\over
4m_hm_j}f_{ahj}.\eqno(3.21)$$ Using all these results in the equation
$\sim\d_\mu A_a^\mu\Phi_h u_j$ we arrive at $$f^3_{ahj}={m_j^2-m_h^2+m_a^2\over
2m_j}f_{ahj},\eqno(3.22)$$ and then from $u_h\Phi_j\Phi_k$ we obtain
$$f^4_{hjk}=0.\eqno(3.23)$$ We have succeeded in expressing all couplings so far
by $f_{abc}$. With these results all remaining terms without Higgs couplings
cancel.

We next turn to the Higgs couplings. From $A_a^\mu u_b\d_\mu\fii_p$ we find
$$f^7_{abp}=m_bf^5_{abp},\quad f^7_{abp}=0\quad{\rm for} \quad a>r\quad {\rm
or}\quad b>r,\eqno(3.24)$$ and from $\d_\mu A_a^\mu u_h\fii_p$ we get
$$f^8_{abp}=-m_bf^5_{abp}\eqno(3.25)$$ and =0 for $b>r$. Finally the terms $\sim
u_a\Phi_h\fii_p$ give $$f^9_{ahp}=-{\mu_p^2\over 2m_a}f^5_{ahp},\quad a\le
r\eqno(3.26)$$ and zero for $a>r$. The terms $\sim u_a\fii_p\fii_q$ lead to
$$f^{10}_{apq}={\mu_q^2-\mu_p^2\over m_a^2}f^6_{apq},\quad a\le r \eqno(3.27)$$
and zero for $a>r$. We see that the Higgs couplings are not completely fixed by
first order gauge invariance. So far the Higgs couplings could be set equal to
zero, but then we would find a breakdown of gauge invariance at second order.

\vskip 1cm \section{Second order gauge invariance} \vskip 0.5cm

Following the inductive construction of Epstein and Glaser \cite{2} in the case
of $T_2$, we have first to calculate the causal distribution
$$D_2(x,y)=T_1(x)T_1(y)-T_1(y)T_1(x).\eqno(4.1)$$ It has a causal support
($\subset \{(x-y)^2\ge 0\}$) and must be decomposed into a retarded and advanced
part: $D_2=R_2-A_2$, ${\rm supp} R_2 \subset V^+$, ${\rm supp} A_2 \subset V^-$.
For diagrams with singular order $\omega\ge 0$ \cite{16} this distribution
splitting is not unique. There are undetermined local terms
$$\sum_{|a|=0}^\omega C_aD^a\delta(x-y):O(x,y):\quad a=(a_\mu)$$ in $R_2$ which
are called normalization terms (or finite renormalization terms in the old
terminology).  $D^a=\prod_\mu\d_{x_\mu}^{a_\mu} $ is a partial differential
operator and $:O(x,y):$ is a Wick monomial.  Finally, we obtain $T_2=R_2-R'_2$,
where $R'_2(x,y)\=d -T_1(x)T_1(y)$.

The main problem is whether gauge invariance can be preserved in the
distribution splitting.  Obviously, $D_2$ (4.1) is gauge invariant:
$$d_QD_2(x,y)=[d_QT_1(x),T_2(y)]+[T_1(x),d_QT_2(y)]=$$
$$=i\d_\mu^x[T_{1/1}^\mu(x),T_1(y)]+i\d_\mu^y[T_1(x),T_{1/1}^\mu(y)] \=d
i\d_\mu^xD^\mu_{2/1}(x,y)+i\d_\mu^yD^\mu_{2/2}(x,y).\eqno(4.2)$$ Since the
retarded part $R_2$ agrees with $D_2$ on the forward light cone $x\in
x+(V^+\setminus \{0\})$ and similarly for $R^\mu_{2/1}, R^\mu_{2/2}$, gauge
invariance of $R_2$ can only be violated by local terms $\sim D^a\delta(x-y)$.
But such local terms also appear as normalization terms in the distribution
splitting if the singular order is $\ge 0$. If the normalization terms $N_2,
N_{2/1}^\mu, N_{2/2}^\mu$ can be chosen in such a way that
$$d_Q(R_2+N_2)=\d_\mu^x(R_{2/1}^\mu+N_{2/1}^\mu)+\d_\mu^y(R_{2/2}^\mu+
N_{2/2}^\mu)\eqno(4.3)$$ holds, then the theory is gauge invariant to second
order. Note that the distribution $T_2=R_2+N_2-R'_2$ then fulfils (4.3), too,
because $R'_2$ is clearly gauge invariant for the same reason as in (4.2). The
local terms on the right-hand side of (4.3), which come from the causal
splitting, are called "anomalies". The ordinary axial anomalies are of the same
kind, they appear in the third order triangle diagrams with axial vector
couplings to fermions (see \cite{6}, Sect.4). The difference is that the axial
anomalies cannot be removed by finite renormalizations.

To prove (4.3) we only have to consider its local part. We concentrate on the
tree graphs because gauge invariance is not a serious problem for second order
loop graphs. Let $R_2$ be the splitting solution of $D_2$ obtained by replacing
$D_m(x-y)$ by $D^{\rm ret}_m(x-y)$. Since $d_Q$ operates only on the field
operators, the local part on the left-hand side of (4.3) is only due to
$d_QN_2$. To calculate the anomalies on the right-hand side of (4.3) we start
from $$D^\mu_{2/1}\=d [T^\mu_{1/1}(x), T_1(y)].\eqno(4.4)$$ The anomalies come
from those terms in $T^\mu_{1/1}$ (3.17) which contain a derivative $\d^\mu$.
These are the second and third term in (3.17.1), the first term in (3.17.2), the
first two in (3.17.4) and the first in (3.17.5). We shall abbreviate these terms
by 17.1/2...  17.5/1 in the following. Commuting the factors with derivative
$\d^\mu$ in these terms with all terms in $T_1(y)$ (2.9-20) we get tree-graph
contributions with four external legs (sectors) which we now have to examine.

\vskip 0.5cm \subsection{Sector $uA\tilde uu$:} \vskip 0.5cm These field
operators come out if we commute the second term in (3.17.1) with the second one
in (2.9) $$(17.1/2)-(2.9/2)=if_{abc}f_{def}A_{\nu a}u_b[\d^\mu A_c^\nu(x),
A_{\lambda d}(y)]u_e\d^\lambda\tilde u_f,\eqno(4.5)$$ where we set the coupling
constant $g=1$ from now on. This gives a result $\sim\d_x^\mu D(x-y)$. After
splitting this causal distribution we get the retarded part $\d^\mu_x D_{\rm
ret} (x-y)$. If now the derivative $\d_\mu^x$ of (4.2) is applied
$$\d_\mu\d_x^\mu D_{\rm ret}(x-y)=-m^2D_{\rm ret}+\delta(x-y)\eqno(4.6)$$ we get
a local term $$A_1=-f_{abc}f_{cef}A_{\nu a}u_bu_e\d_\nu\tilde u_f
\delta(x-y)\eqno(4.7)$$ which is the anomaly. The second term in (4.2) with $x$
and $y$ interchanged gives the same contribution so that we notice the short
rule $$\d^\mu D(x-y)\lra 2\delta(x-y)\eqno(4.8)$$ for the following. Proceeding
in the same way with the third term in (3.17.1) commuted with the second one in
(2.9) we get $$(17.1/3)-(2.9/2)=f_{abc}f_{dcf}u_au_b\d^\nu\tilde u_fA_{\nu d}
\delta(x-y).\eqno(4.9)$$

There are no further contributions in this sector so that (4.7) must cancel
against (4.9) in order to have gauge invariance. We interchange the indices of
summation $b$ and $e$ in (4.7)
$$2A_1=(-f_{abc}f_{cef}+f_{aec}f_{cbf})u_bu_e\d^\nu\tilde u_fA_{\nu a}$$ and add
(4.9), then the total anomaly becomes
$$(-f_{abc}f_{cef}+f_{aec}f_{cbf}-f_{ebc}f_{acf})u_bu_e\d^\nu\tilde u_fA_{\nu
a}.\eqno(4.10)$$ Taking the total asymmetry of $f_{abc}$ into account the
bracket vanishes iff the Jacobi identity is satisfied

\vskip 0.5cm \subsection{Sector $uAAA$:} \vskip 0.5cm As the foregoing one this
is a pure Yang-Mills sector. From the commutator between (3.17.1/2) and (2.9/1)
we get three contributions $$(17.1/2)-(2.9/1)=f_{abc}A_{\nu
a}u_b(x)\Bigl\{f_{cef}A_{\alpha e} \d^\alpha A^\nu_f\d^\mu D+f_{dcf}A_{\lambda
d}\d^\nu A^\lambda_f \d^\mu D\eqno(4.11)$$ $$+f_{dec}A^\nu_d(y)A_{\alpha
e}\d_x^\mu\d_y^\alpha D(x-y)\Bigl\}.  $$

Here in the last term we have a new situation because the distribution
$\d^\mu\d^\alpha D$ has singular order 0. Consequently its retarded part
$\d^\mu\d^\alpha D^{\rm ret}+\alpha_1 g^{\mu\alpha} \delta$ contains a free
normalization term which is part of $\d_\mu N^\mu_{2/1}$ in (4.3), $\alpha_1$ is
a free parameter. Applying the external derivative $\d_\mu$ we get local terms
of the following form
$$G(x)F(y)\d_y^\alpha\delta(x-y)+\alpha_1\d_y^\alpha[G(x)F(y)\delta].
\eqno(4.12)$$ In the $\d\delta$-term we use the identity
$$F(x)G(y)\d^\alpha_x\delta(x-y)+F(y)G(x)\d^\alpha_y\delta(x-y)=$$
$$=F(x)(\d^\alpha G)(x)\delta(x-y)-(\d^\alpha F)(x)G(x)\delta(x-y).\eqno(4.13)$$
Here we have added the other anomaly with $x$ and $y$ interchanged which comes
from $\d_\mu R^\mu_{2/2}$. Similarly, the normalization term of divergence form
in (4.12) will be transformed with help of the relation
$$\d^\alpha_x[F(x)G(y)\delta(x-y)]+\d^\alpha_y[F(y)G(x)\delta(x-y)]=$$
$$=(\d^\alpha F)(x)G(x)\delta(x-y)+F(x)(\d^\alpha G)(x)\delta(x-y).
\eqno(4.14)$$ Summing up, we have the following short rule for the calculation
of this type of local terms:
$$G(x)F(y)\d^\mu_x\d^\alpha_yD(x-y)\lra\Bigl[(\alpha_1+1)\d^\alpha GF
+(\alpha_1-1)G\d^\alpha F\Bigl]\delta(x-y).\eqno(4.15)$$ Using this in (4.11) we
get the following total result for the local terms $$=f_{abc}A_{\nu
a}u_b(x)\Bigl\{f_{cef}A_{\alpha e} \d^\alpha A^\nu_f+f_{dcf}A_{\lambda d}\d^\nu
A^\lambda_f\Bigl\} 2\delta$$ $$+f_{abc}f_{dec}\Bigl\{(\alpha_1+1)(\d_\alpha
u_bA_{\nu a}+u_b \d_\alpha A_{\nu a})A^\nu_dA^\alpha_e$$
$$+(\alpha_1-1)u_bA_{\nu a}(\d_\alpha A^\nu_dA^\alpha_e+ A^\nu_d\d_\alpha
A^\alpha_e)\Bigl\}\delta .\eqno(4.16)$$

From the vanishing of the term $\sim u_bA_{\nu a}A^\nu_d\d_\alpha A^\alpha_e$ we
conclude $$(\alpha_1-1)f_{abc}f_{dec}=0,\eqno(4.17)$$ which implies
$\alpha_1=1$. Then the terms $\sim u_bA^\alpha_e \d_\alpha A^\nu_fA_{\nu a}$
cancel due to the Jacobi identity.  But the term $\sim\d_\alpha u_bA_{\nu
a}A^\nu_dA^\alpha_e$ does not vanish
$$(\alpha_1+1)f_{abc}f_{dec}=4\beta_1.\eqno(4.18)$$ Here a normalization term
$N_2$ in (4.3) is necessary. In fact, the 4-boson coupling $$N_1=-i\beta_1A_{\nu
a}A^\nu_dA_{\alpha b}A^\alpha_e\delta(x-y) \eqno(4.19)$$ with the gauge
variation $$d_QN_1=4\beta_1\d_\alpha u_bA^\alpha_eA_{\nu a}A^\nu_d\delta
\eqno(4.20)$$ gives just the desired local term. Such a normalization term
(4.19) is indeed possible because the first term in (2.9) commuted with itself
gives the following second order tree graph contribution
$$D_2=-f_{abc}f_{def}A_{\mu a}A_{\nu a}[\d^\nu A^\mu_c(x), \d^\alpha
A^\lambda_f(y)]A_{\lambda_d}A_{\alpha e}.$$ The commutator $\sim\d^\nu\d^\alpha
D(x-y)$ has singular order 0 again, which allows the normalization term (4.19).
$\alpha_1=1$ in (4.18) fixes $\beta_1$: $$N_1=-{i\over 2}f_{abc}f_{dec}A_{\nu
a}A^\nu_dA_{\alpha b}A^\alpha_e\delta(x-y). \eqno(4.21)$$ This is the mechanism
how additional couplings are generated by gauge invariance. Note that in (4.17)
no normalization term is possible.

For later use we list the form of all possible normalization terms. They come
from second order tree graphs with two derivatives on the inner line:
$$(2.9)-(2.9): AAAA, (2.10)-(2.10): A\Phi A\Phi, (2.10)-(2.14): A\Phi A \fii,$$
$$(2.14)-(2.14): A\Phi A\Phi,\> A\fii A\fii, (2.14)-(2.15): A\Phi A\fii,
(2.15)-(2.15): A\fii A\fii.\eqno(4.22)$$ In addition we shall need three further
normalization terms $$\Phi\Phi\Phi\Phi, \Phi\Phi\fii\fii, \fii\fii\fii\fii.$$
They are produced by fourth order box diagrams with all derivatives on inner
lines.

\vskip 0.5cm \subsection{Sector $uA\Phi\fii$:} \vskip 0.5cm Now we have the
tools to discuss all cases of compensation of local terms. For
$u_a\fii_qA^\nu_d\d_\nu\Phi_k$ we find the relation
$$2(\alpha_2+1)f^1_{akj}f^5_{djq}-2f_{dac}f^5_{ckq}-2(\alpha_3+1)
f^5_{akp}f^6_{dqp}$$
$$-2(\alpha_1-3)f^5_{ajq}f^1_{djk}-2(\alpha_4-3)f^6_{aqp}f^5_{dkp}=0.
\eqno(4.23)$$ For $u_a\d_\nu\fii_qA^\nu_d\Phi_k$ we have
$$2(\alpha_2-3)f^1_{akj}f^5_{djq}+2f_{dac}f^5_{ckq}-2(\alpha_3-3)
f^5_{akp}f^6_{dqp}$$
$$-2(\alpha_1+1)f^5_{ajq}f^1_{djk}-2(\alpha_4+1)f^6_{aqp}f^5_{dkp}
=0.\eqno(4.24)$$ For $\d_\nu u_a\fii_qA^\nu_d\Phi_k$ we find
$$2(\alpha_2+1)f^1_{akj}f^5_{djq}-2(\alpha_3+1)
f^5_{akp}f^6_{dqp}-2(\alpha_1+1)f^5_{ajq}f^1_{djk}$$
$$-2(\alpha_4+1)f^6_{aqp}f^5_{dkp}=\beta_2.\eqno(4.25)$$ For $u_a\fii_q\d_\nu
A^\nu_d\Phi_k$ we get $$2(\alpha_2-1)f^1_{akj}f^5_{djq}-2(\alpha_3-1)
f^5_{akp}f^6_{dqp}-2(\alpha_1-1)f^5_{ajq}f^1_{djk}$$
$$-2(\alpha_4-1)f^6_{aqp}f^5_{dkp}=0.\eqno(4.26)$$ In choosing different
parameters $\alpha_1,\ldots\alpha_4$ we have split every tree graph contribution
separately. If we sum the terms with the same field operators before splitting,
we have one $\alpha$ only, but the results remain the same as we are now going
to show.

Subtracting (4.24) from (4.23) and (4.26) from (4.23) and subtracting the two
resulting equations we obtain $$f^6_{aqp}f^5_{dkp}=0.$$ Here the sum goes over
$p=1,\ldots t$ and can be regarded as a scalar product of two vectors in ${\bf
R^t}$. We will see below (see (4.44)) that the vectors $(f^5_{dk})_p$ are
non-zero. We make the weak assumption that there are $t$ linear independent
vectors $(f^5_{dk})_p$ for different $d,k$, then it follows $f^6=0$.
Subtracting (4.23) from (4.25) we conclude
$$\beta_2(a,d,k,q)=2f_{dac}f^5_{ckq}-8f^5_{ajq}f^1_{djk}.\eqno(4.27)$$ This will
be simplified below if we have more information about $f^5$. $\beta_2$ belongs
to the normalization term $$N_2=-{i\over 2}\beta_2(a,d,k,q)A_{\nu
a}A^\nu_d\Phi_k \fii_q\delta\eqno(4.28)$$ with $$d_QN_2=\beta_2\d_\nu
u_aA^\nu_d\Phi_k\fii_q\delta+\beta_2 {m_k\over 2}u_kA_{\nu
a}A^\nu_d\fii_q\delta.\eqno(4.29)$$ The last term herein couples this sector to
the sector $uAA\fii$.

\vskip 0.5cm \subsection{Sector $uu\tilde u\fii$:} \vskip 0.5cm In this sector
we have only one combination of external legs, namely $u_au_b\tilde u_d\fii_p$.
The corresponding relation is
$$2f_{abc}f^8_{dcp}-2f^5_{ajp}f^3_{dbj}+2f^5_{bjp}f^3_{daj}+4f^6_{apq}
f^8_{dbq}=0.$$ The origin of the terms is clear from the upper indices.  Since
$f^6=0$ we have $$f^3_{dbj}f^5_{ajp}-f^3_{daj}f^5_{bjp}+m_jf_{abj}f^5_{djp}=0,
\eqno(4.30)$$ We specialize to $d=a$ and insert (3.22):
$$\sum_{j=1}^r{3m_j^2-m_b^2+m_a^2\over 2m_j}\, f_{abj}f^5_{ajp}=0.
\eqno(4.31)$$ If we write a summation symbol then only the indicated index is
summed over. For $a,b,j$ all different, $f_{abj}$ defines a non-singular matrix
and the mass-dependent factor does not alter that. Consequently $f^5$ vanishes
for different indices, only $f^5 _{jjp}$, $j=1,\ldots r$ are different from 0.
That means the Higgs couplings are diagonal, in contrast to the couplings of the
unphysical scalars which are non-diagonal.

Now (4.30) can be simplified
$$f^3_{dba}f^5_{aap}-f^3_{dab}f^5_{bbp}+m_df_{abd}f^5_{ddp}=0 \eqno(4.32)$$
without summation. Interchanging $a$ with $d$ and $b$ with $d$, we get a
homogeneous linear system for $f^5$ where $p$ is a dummy index
$$m_af_{dba}f^5_{aap}-f^3_{adb}f^5_{bbp}+f^3_{abd}f^5_{ddp}=0 \eqno(4.33)$$
$$f^3_{bda}f^5_{aap}+m_bf_{adb}f^5_{bbp}-f^3_{bad}f^5_{ddp}=0.  \eqno(4.34)$$
Using (3.22) it is easy to check that the $3\times 3$ determinant vanishes so
that we get a non-trivial solution. The latter is very simple
$$f^5_{aap}={m_a\over m_d}f^5_{ddp},\eqno(4.35)$$ in particular, $f^5_{aap}=0$
for $a>r$.

With help of (4.35) we can simplify the previous result (4.27) for the
normalization factor $$\beta_2(a,d,k,q)=2{m_d^2-m_a^2\over m_d
m_k}\,f_{dak}f^5_{ddq}= 2{m_d^2-m_a^2\over
m_k^2}\,f_{kda}f^5_{kkq}.\eqno(4.36)$$ By (4.35) this is symmetric in $a, d$ as
it must be (4.28). Furthermore, by means of (4.35) it is easy to check that all
remaining relations in the sector $uA\Phi\fii$ are satisfied. We have still to
show that $f^5\ne 0$. This follows from the following sector.

\vskip 0.5cm \subsection{Sector $uA\Phi\Phi$:} \vskip 0.5cm From
$u_a\d_\nu\Phi_j\Phi_hA^\nu_d$ we get
$$4f_{dac}f^1_{chj}-4(\alpha_1-3)f^1_{ahk}f^1_{djk}-4(\alpha_1+1)
f^1_{ajk}f^1_{dhk}$$
$$-(\alpha_2+1)f^5_{ajp}f^5_{dhp}-(\alpha_2-3)f^5_{ahp}f^5_{djp}
=0,\eqno(4.37)$$ and, assuming $j\ne h$, $u_a\Phi_j\Phi_h\d_\nu A^\nu_d$ gives
$$(\alpha_2-1)(f^5_{ajp}f^5_{dhp}+f^5_{ahp}f^5_{djp})+4(\alpha_1
-1)(f^1_{ajk}f^1_{dhk}+f^1_{ahk}f^1_{djk})=0.\eqno(4.38)$$ Finally $\d_\nu
u_a\Phi_j\Phi_hA^\nu_d$ gives
$$-(\alpha_2+1)(f^5_{ajp}f^5_{dhp}+f^5_{ahp}f^5_{djp})-
4(\alpha_1+1)(f^1_{ajk}f^1_{dhk}+f^1_{ahk}f^1_{djk})=2\beta_3, \eqno(4.39)$$
with $$N_3=-{i\over 2}\beta_3(a,d,j,h)A_{\nu a}A^\nu_d\Phi_j\Phi_h
\delta\eqno(4.40)$$ $$\d_QN_3=\beta_3\d_\nu
u_aA^\nu_d\Phi_j\Phi_h\delta+\beta_3m_j u_jA_{\nu
a}A^\nu_d\Phi_h\delta.\eqno(4.41)$$ Subtracting (4.37) from (4.39) we find
$$\beta_3(a,d,j,h)=-2f_{dac}f^1_{chj}-8f^1_{ahk}f^1_{djk}
-2f^5_{ahp}f^5_{djp}\eqno(4.42)$$ where the first term does not contribute to
(4.40). The result (4.42) remains valid for $j=h$.

Subtracting now (4.37) and (4.38) and using previous results it follows
$$f^5_{ajp}f^5_{dhp}-f^5_{ahp}f^5_{djp}={m^2_j+m^2_h-m^2_c\over 2m_hm_j}
f_{dac}f_{chj}$$ $$-{m^2_k+m^2_j-m^2_a\over
m_jm_k}f_{ajk}{m^2_k+m^2_h-m^2_d\over 4 m_hm_k}f_{dhk}$$
$$+{m^2_k+m^2_h-m^2_a\over m_hm_k}f_{ahk}{m^2_k+m^2_j-m^2_d\over 4
m_jm_k}f_{djk}.\eqno(4.43)$$ In the special case $a=j$ and $d=h$ ($j\ne h$) we
have $$\sum_{p=1}^t f^5_{jjp}f^5_{jjp}={1\over 2m_h^2}\biggl\{\sum_{c=1}
^{r+s}(m_j^2+m_h^2-m_c^2)f_{jhc}f_{jhc}$$
$$-\sum_{k=1}^r{m_k^4-(m_j^2-m_h^2)^2\over2m_k^2}\,f_{jhk}f_{jhk}
\biggl\}.\eqno(4.44)$$ The r.h.s. is known and generally different from 0,
consequently $f^5$ must be also different from 0. In case of only one Higgs
field $t=1$, the Higgs coupling $f^5$ can be calculated from (4.44) as a square
root. For $t>1$ the Higgs couplings are no longer uniquely determined by gauge
invariance. For fixed $j$ equation (4.44) holds for all $h\ne j$ and gives the
same value on the l.h.s. This implies relations between the masses and the
Yang-Mills couplings (see sect.6).

\vskip 0.5cm \subsection{Sector $uAA\Phi$:} \vskip 0.5cm In this sector there is
only one Wick monomial $u_aA_{\nu b}A^\nu_c\Phi_h$ which for $b\ne c$ gives the
relation $$4(f_{bad}f^2_{dch}+f_{cad}f^2_{dbh})-4(f^1_{ahj}f^2_{bcj}
+f^1_{ahj}f^2_{cbj})$$ $$-2(f^5_{ahp}f^7_{bcp}+f^5_{ahp}f^7_{cbp})=m_a\Bigl(
\beta_3(b,c,a,h)+\beta_3(c,b,a,h)\Bigl),\eqno(4.45)$$ where (4.41) has been
taken into account. Substituting (4.42) and previous results we obtain
$$2m_a(f^5_{bhp}f^5_{cap}+f^5_{chp}f^5_{bap})={2\over m_h}\Bigl[
(m_d^2-m_c^2)f_{bad}f_{chd}+(m_d^2-m_b^2)f_{cad}f_{bhd}\Bigl]$$
$$+{m_j^2+m_h^2-m_a^2\over m_hm_j^2}(m_c^2-m_b^2)f_{ahj}f_{bcj}
-{m_j^2+m_h^2-m_b^2\over 2m_hm_j^2}(m_j^2+m_a^2-m_c^2)f_{bhj} f_{caj}$$
$$-{m_j^2+m_h^2-m_c^2\over 2m_hm_j^2}(m_j^2+m_a^2-m_b^2)
f_{chj}f_{baj}.\eqno(4.46)$$ In the case $h=b\ne c$ this leads to
$$m_am_bf^5_{aap}f^5_{bbp}\delta_{ac}=-m_c^2\sum_{d>r}f_{abd} f_{bcd}$$
$$+\sum_{j=1}^r{f_{abj}f_{bcj}\over 4m_j^2}\Bigl[(m_j^2-m_b^2)
(3m_j^2-m_a^2+m_b^2)-m_c^2(m_j^2+m_a^2-m_b^2)\Bigl].\eqno(4.47)$$ For $b\ne h\ne
c$ we find $$\sum_{d>r}(m_c^2f_{bad}f_{chd}+m^2_bf_{cad}f_{bhd})=$$
$$=\sum_{j=1}^r{1\over 4m_j^2}\Bigl\{f_{bhj}f_{caj}[(m_j^2-m_b^2)(3m_j^2-m_a^2+
m_c^2)-m_h^2(m_j^2+m_a^2-m_c^2)]$$
$$+f_{baj}f_{chj}[(m_j^2-m_c^2)(3m_j^2-m_a^2+m_b^2)-m_h^2( m_j^2+m_a^2-m_b^2)]$$
$$+2f_{bcj}f_{ahj}(m_j^2+m_h^2-m_a^2)(m_b^2-m_c^2)\Bigl\}.  \eqno(4.48)$$

In the remainig case $b=c$ we have
$$2m_af^5_{bap}f^5_{bhp}-2m_bf^5_{ahp}f^5_{bbp}=$$
$$=-{(m_k^2+m_a^2-m_b^2)(m_k^2+m_h^2-m_b^2)\over 2m_h m_k^2}f_{bak}f_{bhk}
-2{m_b^2-m_d^2\over m_h}f_{bad}f_{dbh}.\eqno(4.49)$$ For $a=h\ne b=c$ this gives
$$m_bf_{aap}^5f_{bbp}^5={m_b^2\over m_a}\sum_{d>r}(f_{bad})^2$$
$$+\sum_k{(f_{bak})^2\over 4m_am_k^2}\Bigl[(m_k^2-m_b^2)
(-3m_k^2-m_b^2+2m_a^2)+m_a^4\Bigl]\eqno(4.50)$$ and for $h\ne a\ne b=c\ne h$ we
get $$m_b^2\sum_{d>r}f_{bad}f_{bhd}=-\sum_{k=1}^rf_{bak}f_{bhk} {1\over
4m_k^2}$$ $$\times\Bigl[(m_k^2-m_b^2)(-3m_k^2+m_a^2-m_b^2+m_h^2)
+m_a^2m_h^2\Bigl].\eqno(4.51)$$

\vskip 0.5cm \subsection{Sector $uA\fii\fii$:} \vskip 0.5cm From
$u_a\fii_p\d_\nu\fii_qA^\nu_b$ we get
$$4f_{bac}f^6_{cpq}-(\alpha_1+1)f^5_{ajq}f^5_{bjp}-(\alpha_1-3)
f^5_{ajp}f^5_{bjq}$$
$$-4(\alpha_2+1)f^6_{aqv}f^6_{bpv}-4(\alpha_2-3)f^6_{apv}f^6_{bqv}
=0,\eqno(4.52)$$ and, assuming $p\ne q$, $u_a\fii_p\fii_q\d_\nu A^\nu_b$ gives
$$(\alpha_1-1)(f^5_{ajp}f^5_{bjq}+f^5_{ajq}f^5_{bjp})+4(\alpha_2
-1)(f^6_{aqv}f^6_{bpv}+f^6_{apv}f^6_{bqv})=0.\eqno(4.53)$$ Finally $\d_\nu
u_a\fii_p\fii_qA^\nu_b$ gives
$$-(\alpha_1+1)(f^5_{ajp}f^5_{bjq}+f^5_{ajq}f^5_{bjp})-
4(\alpha_2+1)(f^6_{aqv}f^6_{bpv}+f^6_{apv}f^6_{bqv})=2\beta_4, \eqno(4.54)$$
with $$N_4=-{i\over 2}\beta_4(a,b,p,q)A_{\nu a}A^\nu_b\fii_p\fii_q
\delta.\eqno(4.55)$$ Adding (4.53) and (4.54) and using previous results we get
$$\beta_4(a,b,p,q)=-2f^5_{aap}f^5_{aaq}\delta_{ab}\eqno(4.56)$$ where no
summation is involved. The same result remains valid for $p=q$. One easily
checks that all other relations in this sector are fulfilled.  \vskip 0.5cm
\subsection{Remaining sectors:} \vskip 0.5cm In the sector $uAA\fii$ we get
another expression for the normalization factor $\beta_2$ in (4.29) which is
consistent with (4.36). The sector $u\Phi\Phi\fii$ vanishes identically because
$f^4=f^6=f^{10}=0$.  In the sector $u\tilde uu\Phi$ we obtain the relation
$$m_af^5_{bjp}f^5_{dap}-m_bf^5_{ajp}f^5_{dbp}={m_j^2-m_c^2+m_d^2 \over
2m_j}f_{abc}f_{dcj}$$ $$+f_{ajk}f_{dbk}{m_k^2+m_j^2-m_a^2\over
4m_jm_k^2}(m_k^2-m_b^2+m_d^2) -f_{bjk}f_{dak}{m_k^2+m_j^2-m_b^2\over
4m_jm_k^2}(m_k^2-m_a^2+m_d^2).  \eqno(4.57)$$ The sector $u\fii\fii\fii$
vanishes identically. In the sector $u\Phi\Phi\Phi$ we find the following
normalization term $$N_5=-{i\over
2}\beta_5(l,j)\Phi_l^2\Phi_j^2\delta\eqno(4.58)$$ with
$$\beta_5(l,j)=\sum_p{\mu_p^2\over 2m_lm_j}f_{llp}^5f_{jjp}^5.  \eqno(4.59)$$ By
(4.35) this is independent of $l,j$: $$\beta_5=\sum_p{\mu_p^2\over
2}\Bigl({f^5_{aap}\over m_a}\Bigl)^2 \eqno(4.60)$$ with $a<r$ arbitrary.
Finally, in the sector $u\Phi\fii\fii$ we obtain another normalization term
$$N_6=-{i\over 2}\beta_6(h,p,q)\Phi_h^2\fii_p\fii_q\delta\eqno(4.61)$$ where
$$\beta_6(h,p,q)={1\over m_h}\Bigl[-f^5_{hhp}f^5_{hhq} {\mu_p^2+\mu_q^2\over
m_h}-6\sum_u f_{hhu}^5f_{pqu}^{11}\Bigl]$$
$$=-(\mu_p^2+\mu_q^2){f^5_{aap}f^5_{aaq}\over m_a^2}-6\sum_u {f^5_{aau}\over
m_a}f^{11}_{pqu}\eqno(4.62)$$ is independent of $h$.  The pure Higgs coupling
$f^{11}$ (2.20) is still completely free, it will be restricted at third order.
In addition we shall need a pure Higgs normalization term of the form
$$N_7=-{i\over 2}\beta_7(p,q,u,v)\fii_p\fii_q\fii_u\fii_v.  \eqno(4.63)$$ \vskip
1cm \section{Third order gauge invariance} \vskip 0.5cm Instead of (4.4) we now
have to look for local terms $\sim\delta^8 (x-z,y-z)$ in
$$D^\mu_{3/1}(x,y,z)=[T^\mu_{1/1}(x),T_2(y,z)]+[T_1(y),T^\mu_{2/1}(x,z)]
+[T_1(z),\tilde T^\mu_{2/1}(x,y)]\eqno(5.1)$$
$$D^\mu_{3/2}(x,y,z)=[T^\mu_{1/1}(y),T_2(x,z)]+[T_1(x),T^\mu_{2/1}(y,z)]
+[T_1(z),\tilde T^\mu_{2/2}(x,y)]\eqno(5.2)$$
$$D^\mu_{3/3}(x,y,z)=[T_1(x),T^\mu_{2,2}(y,z)]+[T_1(y),T^\mu_{2/2}(x,z)]
+[T^\mu_{1/1}(z),\tilde T_2(x,y)]\eqno(5.3)$$ where $\tilde T_2$ refers to the
inverse $S$-matrix \cite{16}. The first term in (5.1) produces a local term if
the second term in (3.17.1) is commuted with the second order normalization term
(4.21). The latter contains $\delta(y-z)$ and the commutator $\sim\d^\mu D(x-y)$
gives another $\delta(x-y)$ by the usual mechanism (4.8). The result is
$$(17.1/2)-(4.21)=-2f_{abc}f_{cb'e}f_{a'd'e}u_bA_{\nu a}A^\nu_{a'} A_{\lambda
b'}A^\lambda_{d'}\delta(x-y)\delta(y-z).\eqno(5.4)$$

To examine the second term in (5.1) we use the fact that $\tilde
T^\mu_{2/1}=-T^\mu_{2/1}+\ldots$ plus terms which give to local contribution.
From (4.19) we have $$\d_\mu^xT^\mu_{2/1}(x,z)\vert_{\rm
loc}=2f_{abc}f_{dec}\d_\lambda u_bA_e^\lambda A_{\nu
a}A^\nu_d\delta(x-z).\eqno(5.5)$$ If this is commuted with the second term in
(2.9), the anti-ghost - ghost contraction has two derivatives so that the
resulting C-number distribution has $\omega=0$ and, after splitting, allows a
normalization term $$(2.9/2)-(5.5)=-2i\alpha
f_{a'b'b}f_{abc}f_{dec}u_{b'}A_{\lambda a'}A^\lambda_eA_{\nu
a}A^\nu_d\delta(y-x)\delta(x-z).\eqno(5.6)$$ After renaming the summation
indices this has the same form as (5.4). However, the contributions from the
second and third member in (5.1) cancel each other and similarly in (5.2). But
in (5.3) these normalization terms survive and after suitable choice of $\alpha$
in (5.6) compensate the anomaly (5.4). Then the sector $uAAAA$ is gauge
invariant. The situation is the same in the other sectors $uAA\Phi\Phi$,
$uAA\Phi\fii$ and $uAA\fii\fii$ containing $A$'s. Here, instead of $N_1$ the
normalization terms $N_2, N_3$ and $N_4$ come into play.

Next we turn to the sector $u\Phi^3\fii$ where we get two anomalies
$$(3.17.4/1)-(4.61)=f^5_{ahp}\beta_6(j,p,u)u_a\Phi_h\Phi_j^2\fii_u$$
$$(3.17.4/2)-(4.58)=-2f^5_{ahp}\beta_5(h,j)u_a\Phi_h\Phi_j^2\fii_p.
\eqno(5.7)$$ They must cancel each other because no normalization term is
possible.  This leads to the relation
$$\sum_{q=1}^tf^5_{aaq}\beta_6(j,q,p)=2f^5_{aap}\beta_5(a,j).  \eqno(5.8)$$ In
case of one physical scalar $(t=1)$ this allows to determine the pure Higgs
coupling $f^{11}$ in $\beta_6$ (4.62).  Similarly, in the sector $u\Phi\fii^3$
we find the relation $$2\sum_{v=1}^t
f^5_{aav}\beta_7(v,p,q,u)=f^5_{aap}\beta_6(a,q,u), \eqno(5.9)$$ which, for
$t=1$, determines the quartic Higgs coupling (4.63). The remaining sectors
$u\Phi^4$ and $u\Phi^2\fii^2$ are automatically gauge invariant due to the facts
that $\beta_5$ (4.60) is constant and $\beta_6(h,p,q)$ (4.62) is independent of
$h$.

It is instructive to discuss the important special case $t=1$ of one physical
scalar in detail. Then (5.8) can be simplified as follows
$$\beta_6(j,1,1)=2\beta_5(a,j),$$ which, by (4.60) and (4.62), leads to
$$f^{11}_{ppp}=-{\mu_p^2\over 2m_a}f^5_{aa1}.\eqno(5.10)$$ Here $\mu_p$ is the
Higgs mass and $a$ is arbitrary. From (5.9) we get $$\beta_7(1,1,1,1)={1\over
2}\beta_6(a,1,1)={\mu_p^2\over 2} \Bigl({f^5_{aa1}\over
m_a}\Bigl)^2.\eqno(5.11)$$

Let us now collect all trilinear purely scalar coupling terms
$$V_1=i\Bigl(f^9_{hj1}\Phi_h\Phi_j\fii+f^{11}_{111}\fii^3 \Bigl)=$$ $$=-{i\over
2}{\mu_p^2\over m_a}f^5_{aa1}\fii\Bigl(\sum_j\Phi_j^2 +\fii^2\Bigl)\eqno(5.12)$$
and the quartic terms $N_5, N_6$ and $N_7$ $$V_2=-{i\over
2}\Bigl(\sum_{lj}\beta_5(l,j)\Phi_l^2\Phi_j^2+
\beta_6\fii^2\sum_j\Phi_j^2+\beta_7\fii^4\Bigl)=$$ $$=-{i\over 2}{\mu_p^2\over
2}\Bigl({f^5_{aa1}\over m_a}\Bigl)^2
\Bigl(\fii^2+\sum_j\Phi_j^2\Bigl)^2.\eqno(5.13)$$ Introducing the coupling
constant $g$ again, we must multiply (5.12) by $g$ and (5.13) by $g^2/2!$
because this is the second order contribution. Then the total scalar potential
is equal to $$V_\fii=-ig^2{\mu_p^2\over 8m_a^2}(f^5_{aa1})^2\Bigl[\Bigl(\fii^2
+\sum_j\Phi_j^2\Bigl)^2+{4m_a\over gf^5_{aa1}}\fii\Bigl(\fii^2
+\sum_j\Phi_j^2\Bigl)\Bigl].\eqno(5.14)$$ Completing the square inside the
square bracket just amounts to addition of a mass term for the Higgs field
$$V(\fii)=V_\fii-{i\over 2}\mu_p^2\fii^2=$$ $$=-ig^2{\mu_p^2\over
8m_a^2}(f^5_{aa1})^2\Bigl[\fii^2+\sum_j \Phi_j^2+{2m_a\over
gf^5_{aa1}}\fii\Bigl]^2.\eqno(5.15)$$ This is the asymmetric Higgs potential. In
fact, introducing the shifted Higgs field $$\tilde\fii=\fii+a,\quad a={m_a\over
gf^5_{aa1}},\eqno(5.16)$$ the Higgs potential (5.15) assumes a symmetric
double-well form $$V\sim\Bigl(\tilde\fii^2+\sum_j\Phi^2-a^2\Bigl)^2.$$ The
shifted Higgs field then has a non-vanishing vacuum expectation value $a$
(5.16), so that we have recovered (i.e. actually deduced) the usual Higgs
mechanism.  \vskip 1cm \section{Derivation of the electroweak gauge theory}
\vskip 0.5cm Let us seek all gauge theories with three massive gauge fields
$m_1, m_2, m_3\ne 0$ and one massless photon field $m_4=0$. There are many
4-dimensional Lie algebras, but we will see that gauge invariance is strong
enough to fix the $f_{abc}$ uniquely.

We put $a=4, d=2, j=1, h=2$ in (4.43) $$0=f_{243}f_{321}{m_1^2+m_2^2-m_3^2\over
2m_1m_2}+f_{423} f_{213}{m_3^2+m_2^2\over m_2m_3}\cdot{m_3^2+m_1^2-m_2^2\over
4m_1m_3}$$ $$={f_{243}f_{321}\over 4m_1m_2m_3^2}(m_3^2m_1^2+2m_3^2m_2^2-
3m_3^4-m_1^2m_2^2+m_2^4).\eqno(6.1)$$ Since the bracket is different from zero,
we must either have $f_{243}=0$ or $f_{321}=0$. We shall verify below that the
second alternative leads to the trivial solution $f=0$ so we concentrate on the
first case. For $a=4, d=1, h=2, j=1$ we find from (4.43)
$$0={f_{143}f_{321}\over 4m_1m_2m_3^2}[m_3^2(m_2^2+2m_1^2-3m_3^2)
+m_1^2(m_1^2-m_2^2)],\eqno(6.2)$$ which implies $f_{143}=0$. Next we put $j=1,
h=2$ in (4.44) and also $j=1, h=3$: $$\sum_p(f_{11p}^5)^2={1\over
2m_2^2}\Bigl[(f_{123})^2(m_1^2+m_2^2 -m_3^2)-$$
$$-(f_{123})^2{m_3^4-(m_1^2-m_2)^2\over 2m_3^2}\Bigl]$$ $$={1\over
2m_3^2}\Bigl[(f_{132})^2(m_1^2+m_3^2-m_2^2)-
(f_{132})^2{m_2^4-(m_1^2-m_3^2)^2\over 2m_2^2}\Bigl].\eqno(6.3)$$ This implies
$$\Bigl({f_{124}\over f_{123}}\Bigl)^2=2{m_3^2(m_3^2-m_1^2)+
m_2^2(m_1^2-m_2^2)\over m_3^2(m_1^2+m_2^2)}.\eqno(6.4)$$ If the r.h.s. is
different from zero we have $f_{124}\ne 0$, otherwise the solution would be
trivial.  Then it follows from (3.19) that $m_1=m_2$ which is the equal mass of
the W-bosons. This simplifies (6.4) as follows $$\Bigl({f_{124}\over
f_{123}}\Bigl)^2={m_3^2\over m_1^2}-1, \eqno(6.5)$$ which implies $m_3>m_1$.
Defining the weak mixing angle $\Theta$ by $${m_1\over
m_3}=\cos\Theta,\eqno(6.6)$$ we have $$\Bigl({f_{124}\over
f_{123}}\Bigl)^2=\tan^2\Theta.$$ Since a common factor in the $f$'s can be
absorbed in the coupling constant $g$, we end up with
$$f_{124}=-\sin\theta,\quad f_{123}=-\cos\Theta\eqno(6.7)$$ in agreement with
the Weinberg-Salam model \cite{8,9}. All other structure constants follow by
asymmetry. The signs in (6.7) have been chosen according to standard convention,
as well as $m_3$ for $m_Z$. Of course any permutation of the indices 1,2,3 is
possible, but the solution remains the same.

It remains to discuss the possibility $f_{123}=0$. Then it follows from (6.3)
that $m_1=m_2=m_3$ and $|f_{124}|=|f_{134}|$. If we now put $a=h=4$ and $d=j=1$
in (4.43) we arrive at $$0={1\over 2}(f_{142}f_{241}+f_{143}f_{341})-{1\over
4}(f_{412} f_{142}+f_{413}f_{143})$$ $$=-{1\over
4}\Bigl((f_{142})^2+(f_{143})^2\Bigl).$$ Hence, all $f$'s vanish in this case.

It is not hard to verify that for the unique non-trivial solution (6.7) all
conditions for gauge invariance are satisfied. By means of (6.7) all couplings
can be calculated in the case of one Higgs field and are in complete agreement
with the standard electroweak theory \cite{13}.  But we could use any number
$t\ge 1$ of Higgs fields. Only for $t=1$ the couplings are completely determined
by gauge invariance.

In the same way one can construct the gauge theory with only two massive gauge
fields $m_1, m_2\ne 0$ and one massless field $m_3=0$. This is not the $SU(2)$
Higgs-Kibble model often discussed in the literature \cite{17} which has three
massive fields. It turns out that $m_1=m_2$ must be equal, so that this theory
is a hypothetical electroweak theory without neutral currents. Therefore, the
gauge principle cannot explain why there are neutral currents in nature.  % % %

\end{document}